\documentclass[12pt]{article}
\def \cn{Collaboration}

\newcommand{\thspace}{\kern.08333em}

\def \beq{\begin{equation}}
\def \eeq{\end{equation}}
\def \bea{\begin{eqnarray}}
\def \eea{\end{eqnarray}}

\newcommand{\bay}{\begin{array}}
\newcommand{\eay}{\end{array}}
\begin{document}
\rightline{UCSD-PTH-00-08}
\rightline{TECHNION-PH-00-27}
\rightline{hep-ph/yymmddd}
\rightline{April 2000}
\bigskip
\bigskip
\centerline{\bf EXTRAPOLATING SU(3) BREAKING EFFECTS}
\centerline{\bf FROM $D$ TO $B$ DECAYS}
\bigskip
\centerline{\it Michael Gronau}
\centerline{\it Physics Department, Technion -- Israel Institute of
Technology}
\centerline{\it 32000 Haifa, Israel}
\medskip
\centerline{and}
\medskip
\centerline{\it Dan Pirjol}
\centerline{\it Department of Physics, University of California at San Diego}
\centerline{\it  La Jolla, CA 92093}
\bigskip
\centerline{\bf ABSTRACT}
\medskip
\begin{quote}
We consider two SU(3) breaking parameters, $R_1(m_B)$ and $R_2(m_B)$,
appearing in a relation between $B^+\to K\pi$ and $B^+\to\pi\pi$ amplitudes,
which plays an important role in determining the weak phase $\gamma$.
We identify an isospin-related quantity $R_2(m_D)$
measured in $D$ decays, exhibiting large SU(3) breaking which is likely
due to nonfactorizable effects. With a cautious remark
about possible nonfactorizable SU(3) breaking in $B$ decays, we proceed
to calculate factorizable SU(3) breaking corrections. Applying heavy quark
symmetry to semileptonic $D$ and $B$ decay form factors, we find that
SU(3) breaking in $R_2(m_B)/R_1(m_B)$ may be significantly larger
than estimated from certain model calculations of form factors.
\end{quote}
\medskip

\leftline{\qquad PACS codes:  12.15.Hh, 12.15.Ji, 13.25.Hw, 14.40.Nd}
\vfill
\newpage

Weak nonleptonic decays of $B$ mesons provide an important source of
information about the elements of the Cabibbo-Kobayashi-Maskawa (CKM) matrix.
Flavor SU(3) symmetry of strong interactions plays an essential
role in some of the methods proposed to determine the weak phases \cite{SU3}.
First order SU(3) breaking effects in hadronioc $B$ decays may be
parametrized in a completely general way in terms of
several unknown parameters \cite{SU3br} some of which can be determined from
experiments. In certain hadronic amplitudes, such as in $B\to \bar D\pi$,
experimental evidence exists for factorization in terms of products of
two current matrix elements \cite{BHP}. In these cases, the corresponding
SU(3)
breaking parameters are given by
ratios of $K$ and $\pi$ decay constants and ratios of $B/B_s$ to $D/D_s$ form
factors. In decays to two charmless pseudoscalar mesons,
which are useful for weak phase determinations \cite{MGrev}, experimental
evidence for factorization of hadronic matrix elements is still lacking.
It was argued recently \cite{BBNS} that within QCD nonfactorizable
corrections due to hard gluon exchange are calculable and those which are
due to soft exchanges
are suppressed by $\Lambda_{\rm QCD}/m_b$ in a heavy quark expansion. Actual
calculations of these corrections, controlling the former in a
model-independent
manner and showing that the latter are indeed small, are both
desirable and challenging. Furthermore, in order to treat SU(3) breaking
within the factorization approximation, one still needs the values of
certain ratios of unmeasured form factors, for which one oftens relies on
theoretical models.

The purpose of this Letter is to learn about SU(3) breaking
in $B$ decays from the corresponding measured effects in $D$ decays.
SU(3) breaking does not necessarily decrease monotonously with the decaying
heavy quark mass. We will try to address the two relevant questions, of
factorizable and nonfactorizable SU(3) violating corrections to hadronic
decays, and of SU(3) breaking in semileptonic form factors which
are used in the factorization approximation.

In general, soft final state interactions which
spoil factorization are expected to affect $D$ and $B$ decays differently.
It was often argued \cite{Dres}, and it has recently been shown by an actual
calculation \cite{MGres}, that $D$ decay amplitudes involve large
contributions
from nearby light $q\bar q$ resonances which induce large SU(3) breaking
effects. Such effects are not expected in $B$
decays. To avoid resonance effects, and thus study $D$ and $B$ decays on
common grounds, we will consider only  decays to ``exotic" final states
involving $\pi\pi$ in $I=2$ and $K\pi$ in $I=3/2$.

We will find very large SU(3) breaking in hadronic $D$ decays, in the
absence of
resonant terms, implying in the most likely scenario large nonfactorizable
corrections. This should serve
as a warning for what may be the case also in $B$ decays. In the factorization
approximation, we then proceed to calculate SU(3) breaking in hadronic $B$
decays, where $B$ meson form factors are obtained from those measured for
$D$ by applying a heavy quark symmetry scaling law. Our result will be compared
with a model-dependent calculation.

As our test case, we consider an SU(3) relation
between the isospin $I=3/2$ amplitude in $B\to K\pi$ and the $I=2$
amplitude in $B\to\pi\pi$ \cite{NR2,MN}
\bea\label{1}
A(B^+\to K^0\pi^+) & + & \sqrt2 A(B^+\to K^+\pi^0) =
\sqrt2\tan\theta_c(R_1 - \delta_+ e^{-i\gamma} R_2)A(B^+\to \pi^0\pi^+)~,
\nonumber\\
\delta_+ & \equiv & -[3/(2\lambda\vert V_{ub}/V_{cb}\vert)][(c_9+
c_{10})/(c_1+c_2)] = 0.66\pm 0.15~.
\eea
This SU(3) relation generalizes a triangle relation proposed in \cite{GLR} by
including, in addition to the current-current (``tree") contributions, also
the
effects of dominant electroweak
penguin (EWP) amplitudes given by the second term on the right-hand-side.
Eq.~(\ref{1}) and its charge-conjugate were proposed  as a way for determining
the weak phase $\gamma \equiv {\rm Arg} V^*_{ub}$. Other suggestions for using
$B\to K\pi$ decays to study $\gamma$ were discussed in \cite{BKpi}.

The coefficients $R_{1,2}$ in Eq.~(\ref{1}) parametrize SU(3) breaking effects
and are in general
complex numbers. In the SU(3) limit they are both equal to 1. Knowledge of
the precise values of $R_1$ and $R_2/R_1$, in the presence of SU(3) breaking,
is crucial for an accurate determination of $\gamma$ \cite{NR2,BF,critical}.
Using the factorization approximation, it is customary to apply the value
$R_1\simeq f_K/f_\pi= 1.22$ to the tree part.
SU(3) breaking corrections to the EWP-to-tree ratio $R_2/R_1$ were
estimated in
the generalized factorization approximation, assuming a certain model-dependent
value for the ratio of $B$ to $K$ and $B$ to $\pi$ form factors, and were
found
to amount to a few percent \cite{NR2,MN}. Numerically, this follows from an
accidental cancellation between the contributions of the color-allowed and
color-suppressed amplitudes. In addition, nonfactorizable SU(3) breaking
corrections can in principle be significant \cite{BF}.

Our main concern will be the SU(3) breaking parameter $R_2$.
We will show that there exists a corresponding quantity $R_2(m_D)$, which has
already been measured in $D$ decays exhibiting large SU(3) breaking.
It will be argued that this effect is likely due to
nonfactorizable corrections. It is not obvious why such effects should be
much suppressed in $B$ decays. With this cautious remark, we
nevertheless
assume factorization in order to calculate $R_1(m_B)$ and $R_2(m_B)$ in this
approximation. Using heavy quark symmetry to extrapolate form factors
from measured semileptonic $D$ decays to $B$ decays, we calculate
factorizable SU(3) breaking in $R_2(m_B)/R_1(m_B)$ and compare with estimates
based on certain models for form factors.

For completeness, and in order to define $R_1$ and $R_2$ in broken SU(3) and
to prove Eq.~(\ref{1}), we
start by quickly reviewing the SU(3) structure of the amplitudes
entering Eq.~(\ref{1}). The tree and electroweak penguin four-quark operators
describing charmless decays transform under flavor SU(3) as a sum of
$\overline{\bf 3},~{\bf 6}$ and $\overline{\bf 15}$~\cite{EWP}
\begin{eqnarray}\label{BHam}
& &{\cal H}^{\Delta S=1}_T +{\cal H}^{\Delta S=0}_T + {\cal H}^{\Delta
S=1}_{EWP} =\\
& &\quad \frac{G_F}{\sqrt2}\lambda_u^{(s)}
[\frac12(c_1-c_2)(-\bar {\bf 3}^{(a)}_{I=0} - {\bf 6}_{I=1}) +
\frac12(c_1+c_2)(-\overline {\bf 15}_{I=1} - \frac{1}{\sqrt2}\overline
{\bf 15}_{I=0}
+\frac{1}{\sqrt2}\bar {\bf 3}^{(s)}_{I=0})] \nonumber\\
&+& \frac{G_F}{\sqrt2} \lambda_u^{(d)}
[\frac12(c_1-c_2)({\bf 6}_{I=\frac12} - \bar {\bf 3}^{(a)}_{I=\frac12}) +
\frac12(c_1+c_2)(-\frac{2}{\sqrt3}\overline {\bf 15}_{I=\frac32} -
\frac{1}{\sqrt6}\overline {\bf 15}_{I=\frac12}
+\frac{1}{\sqrt2}\bar {\bf 3}^{(s)}_{I=\frac12})] \nonumber\\
&-& \frac{G_F}{\sqrt2}
\frac{\lambda_t^{(s)}}{2}\left(
\frac{c_9-c_{10}}{2}(3\cdot {\bf 6}_{I=1} + \bar {\bf 3}^{(a)}_{I=0} ) +
\frac{c_9+c_{10}}{2}( -3\cdot\overline {\bf 15}_{I=1}
-\frac{3}{\sqrt2}\overline {\bf 15}_{I=0}
-\frac{1}{\sqrt2}\bar {\bf 3}^{(s)}_{I=0} )\right)\nonumber~,
\eea
where $\lambda_{q}^{(q')}=V^*_{qb}V_{qq'}$.
The explicit expressions of the four-quark operators appearing in the
Hamiltonian can be found in \cite{EWP}.

The left-hand-side of Eq.~(\ref{1}) receives only contributions from
the $\Delta S=1,~I=1$ terms in the weak Hamiltonian, which transform as
{\bf 6}
and $\overline{\bf 15}$, (QCD penguin operators are pure $I=0$ and do not
contribute)
\bea\nonumber
& &A(B^+\to K^0\pi^+) + \sqrt2 A(B^+\to K^+\pi^0) = \\\label{3}
& &\qquad \lambda_u^{(s)}(
C_{\overline{15}_{I=1}} + C_{6_{I=1}}) + \lambda_t^{(s)}\left(
-\frac32 \frac{c_9+c_{10}}{c_1+c_2} C_{\overline{15}_{I=1}} +
\frac32 \frac{c_9-c_{10}}{c_1-c_2} C_{6_{I=1}}\right)~.
\eea
Here
\beq\label{I=1}
C_{\overline{15}_{I=1}}(m_B) = \frac{G_F}{\sqrt2}\frac12(c_1+c_2)
(\langle  K^0\pi^+|-\overline{\bf 15}_{I=1}|B^+\rangle
 + \sqrt2 \langle K^+\pi^0|-\overline{\bf 15}_{I=1}|B^+\rangle)
\eeq
and
\beq
C_{6_{I=1}}(m_B) = \frac{G_F}{\sqrt2}\frac12(c_1-c_2)
(\langle  K^0\pi^+| -{\bf 6}_{I=1}|B^+\rangle
+ \sqrt2 \langle K^+\pi^0|-{\bf 6}_{I=1}|B^+\rangle)
\eeq
are hadronic matrix elements of operators transforming as $\overline{\bf 15}$
and ${\bf 6}$.
Using the approximate equality
\beq
\frac{c_9+c_{10}}{c_1+c_2} \approx \frac{c_9-c_{10}}{c_1-c_2}\approx
-1.12\alpha~,
\eeq
which holds to better than $3\%$ \cite{BBL}, one finds
\beq
A(B^+\to K^0\pi^+) + \sqrt2 A(B^+\to K^+\pi^0) =
\lambda_u^{(s)}[(C_{\overline{15}_{I=1}} + C_{6_{I=1}}) - \delta_+
e^{-i\gamma}
(C_{\overline{15}_{I=1}} - C_{6_{I=1}})]~.
\eeq
On the other hand, the amplitude on the right-hand-side of (\ref{1}) is given
by the matrix element of the $\Delta S=0,~I=3/2$ term in the weak Hamiltonian,
(we neglect a very small EWP contribution \cite{EWP})
\bea\label{4}
\sqrt2 A(B^+\to \pi^+\pi^0) = \lambda_u^{(d)} C_{\overline{15}_{I=3/2}}~,
\eea
where
\bea\label{I=3/2}
C_{\overline{15}_{I=3/2}}(m_B) &=& \frac{G_F}{\sqrt2}(c_1+c_2)\sqrt\frac23
 \langle \pi^+\pi^0|-\overline{\bf 15}_{I=3/2}|B^+\rangle~.
\eea

Taking the ratio of
(\ref{3}) and (\ref{4}) reproduces the
factor on the right-hand-side of (\ref{1}) with
\bea\label{R12}
R_1(m_B) =
\frac{C_{\overline{15}_{I=1}}+C_{6_{I=1}}}{C_{\overline{15}_{I=3/2}}}~,
\qquad
R_2(m_B) =
\frac{C_{\overline{15}_{I=1}}-C_{6_{I=1}}}{C_{\overline{15}_{I=3/2}}}~.
\eea
Both final states on the left-hand-side of (\ref{3}) and (\ref{4}) belong to
a {\bf 27} multiplet of SU(3), such that the matrix elements of
$\overline{\bf 15}_{I=1}$ and $\overline{\bf 15}_{I=3/2}$ are related in the
SU(3) limit, $C_{\overline{15}_{I=1}}=C_{\overline{15}_{I=3/2}}$.
(The different numerical factors defining these amplitudes in Eqs.~(\ref{I=1})
and (\ref{I=3/2}) are related to the different isospins involved).
Furthermore,
the matrix element of {\bf 6} in (\ref{3}) vanishes in the same limit, such
that $R_1=R_2=1$ in the SU(3) symmetric case.
However, in broken SU(3) $C_{\overline{15}_{I=1}}\neq
C_{\overline{15}_{I=3/2}}$
and $C_{6_{I=1}}\neq 0$, which causes both $R_1$ and $R_2$ to differ from
unity.

Whereas $R_1(m_B)$ and $R_2(m_B)$ are purely theoretical quantities,
which cannot be directly measured, we prove now that another SU(3) breaking
parameter,
\bea\label{2}
R_2(m_D) = -\frac{V_{us}}{V_{ud}}\, \frac{A(D^-\to K^0\pi^-)}{\sqrt2
A(D^-\to \pi^-\pi^0)}~,
\eea
measured in $D$ decays, is related to $R_2(m_B)$ by isospin in a fictitious
heavy quark limit $m_c=m_b$.

The final states in the numerator and denominator of $R_2(m_D)$ have quantum
numbers
$|I=\frac32,~I_3=-\frac32\rangle$
and $|I=2,~,I_3=-1\rangle$, respectively, and belong to the same isospin
multiplets as the states
$|K^0\pi^+\rangle + \sqrt2 |K^+\pi^0\rangle$ and $|\pi^+\pi^0\rangle$ in
(\ref{1}).
The initial states $D^-$ and $B^+$ are related to each other by
isospin in the (fictitious) limit of identical heavy quarks.
The weak Hamiltonian responsible for the relevant $\bar D$ decays is
\bea\label{DHam}
{\cal H}_W &=&
\frac{G_F}{\sqrt2} V_{ud}^* V_{cs}
[\frac12(c_1-c_2)\sqrt2\,\, {\bf 6}_{I=1} -
\frac12(c_1+c_2) \sqrt2\,\, \overline {\bf 15}_{I=1}] \\
&+& \frac{G_F}{\sqrt2}\,  V_{us}^* V_{cs}
[(c_1+c_2)( \frac{1}{\sqrt3} \overline {\bf 15}_{I=3/2} -
\sqrt{\frac23}\,\overline {\bf 15}_{I=1/2}) +
(c_1-c_2)\, {\bf 6}_{I=1/2}]
\nonumber~,
\eea
where we neglect a small CP-violating contribution proportional to
$\frac12(V_{us}^* V_{cs}+
V_{ud}^* V_{cd})={\cal O}(\lambda^5)$ in the Cabibbo-suppressed part
and very small contributions of penguin operators \cite{GoGr}.

The $\Delta S=1$ $(\Delta S=0)$ $I=1,~I_3=-1$ $(I=\frac32,~I_3=-\frac12)$
operators in (\ref{DHam}) are the isospin partners of the $I=1,~I_3=0$
$(I=\frac32,~I_3=\frac12)$
operators in the $B$ decay Hamiltonian (\ref{BHam}). (This can also be
shown in
terms of their quark structure). Therefore, in the limit of identical heavy
quarks, isospin symmetry of strong interactions relates the amplitudes for
$D^-$ decays in (\ref{2}) to those in $B^+$ decays
\bea
A(D^-\to K^0\pi^-) &=& V_{ud}^* V_{cs}(C_{\overline{15}_{I=1}}(m_D) -
C_{6_{I=1}}(m_D))~,\\
\sqrt2 A(D^-\to \pi^-\pi^0) &=& - V_{us}^* V_{cs}
C_{\overline{15}_{I=3/2}}(m_D)~.
\eea
Taking the ratio of these amplitudes yields the SU(3) breaking parameter
$R_2(m_D)$ given in Eq.~(\ref{2}).

The experimental value of the ratio of amplitudes (\ref{2}) is \cite{RPP}
\bea\label{R2number}
|R_2(m_D)| = 0.56 \pm 0.08~.
\eea
The large SU(3) breaking effect in $R_2(m_D)$ is somewhat surprising since the
relevant final states are exotic, $I=\frac32$ and $2$, and thus receive
no resonant contributions \cite{MGres}. The large deviation of the ratio
$|R_2(m_D)|$ from 1 raises the concern of a similar large SU(3) breaking effect
in the $B$ case. In view of this possibility, let us review
previous attempts and difficulties in explaining the numerical value
(\ref{R2number}).

A common way of studying SU(3) breaking in hardonic $D$
(and $B$) decays is by using the generalized factorization approach
\cite{fact}.
In this approach one finds
\bea\label{R2D}
R_2(m_D) = \frac{a_2^{(DK\pi)}}{a_1^{(D\pi\pi)}+a_2^{(D\pi\pi)}}
\,\frac{f_K}{f_\pi}\,
\frac{F_0^{D\pi}(m_K^2)}{F_0^{D\pi}(m_\pi^2)} +
\frac{a_1^{(DK\pi)}}{a_1^{(D\pi\pi)}+a_2^{(D\pi\pi)}}
\,\frac{m_D^2-m_K^2}{m_D^2-m_\pi^2}\,
\frac{F_0^{DK}(m_\pi^2)}{F_0^{D\pi}(m_\pi^2)}\,.
\eea
The phenomenological parameters $a_{1,2}$, describing the external and
internal
$W$-emission amplitudes respectively, are related to corresponding Wilson
coefficients through $a_{1,2}=c_{1,2}+\zeta c_{2,1}$. The parameter $\zeta$
is process- and scale-dependent and is determined from experiments.
When fitting nonleptonic two-body $D\to K\pi$ decays, using
$F_0^{DK}(m_\pi^2)=0.77$ \cite{CLEO} and $F_0^{D\pi}(m_\pi^2)=0.7$ \cite{BSW},
one obtains \cite{fact}
$a_1^{(DK\pi)}=1.26$ and $a_2^{(DK\pi)}=-0.51$, corresponding to
$\zeta(m_c)=0$.
This is compatible with neglecting $1/N_c$ contributions in charm decays
\cite{Nc}. This fit neglects, however, resonance contributions in nonexotic
channels which, when included in an appropriate way, modifies the extracted
values of $a_{1,2}$ to become $a_1^{(DK\pi)}=1.06,~a_2^{(DK\pi)}=-0.64$
\cite{MGres}.

An attempt was made \cite{CC} to explain the large SU(3) breaking in
(\ref{R2number}) by using Eq.~(\ref{R2D}). In this attempt one faces three
kinds of problems. First, there is an uncertainty in the values of
$a_i^{(DK\pi)}$ due to resonance contributions in fitted nonexotic $D$ decays.
Second, the values of $a_i^{(D\pi\pi)}$ may differ from those of
$a_i^{(DK\pi)}$
which causes another uncertainty.
In fact, a determination of $a_i^{(D\pi\pi)}$ from the corresponding Cabibbo
suppressed decays (neglecting resonance contributions) gives very different
results for $a_2$ compared with the $D\to K\pi$ case
\bea
a_1^{(D\pi\pi)}=1.05\left(\frac{0.7}{F_0^{D\pi}(m_\pi^2)}\right)~,\qquad
a_2^{(D\pi\pi)}=-0.07\left(\frac{0.7}{F_0^{D\pi}(m_\pi^2)}\right)~,
\eea
where $F_0^{D\pi}(m_\pi^2)=0.7$ \cite{BSW} is used for normalization.
This large deviation was blamed on inelastic hadronic rescattering \cite{KK}.

Finally, a third uncertainty in evaluating $R_2(m_D)$ using (\ref{R2D}) is
due to the present experimental error in the ratio of form factors
$F_0^{DK}(0)/F_0^{D\pi}(0)$. It was noted in \cite{CC} that the value of
$R_2(m_D)$ is very sensitive to this ratio. In Table 1 we list the results of
four experiments for which the average value is $F_0^{DK}(0)/F_0^{D\pi}(0) =
1.00\pm 0.08$.

\begin{center}
\begin{tabular}{|l|c|c|c|c|}
\hline
 & Mark III \cite{MarkIII} & CLEO \cite{CLEO+}  &
CLEO \cite{CLEO0} & E687 \cite{E687}\\
\hline
\hline
$\frac{F_0^{DK}(0)}{F_0^{D\pi}(0)}$ & $0.951\pm 0.214$ & $1.054\pm 0.246$ &
$0.990\pm 0.230$ & $1.000 \pm 0.110$ \\
\hline
\end{tabular}
\end{center}
\begin{quote} {\bf Table 1.}
Experimental results for the ratio of $D\to \pi(K)$ form factors at $q^2=0$.
In quoting the numbers we used $|V_{cd}/V_{cs}|=0.226$.
\end{quote}

We conclude that it is difficult to evaluate $R_2(m_D)$ and to explain its
experimental value in a reliable manner within the generalized factorization
approach. It is not entirely impossible that the failure to account for
this large SU(3) breaking is due to resonant contributions in other $D$ decay
processes which modify the extracted values of $a_i$. Assuming, for instance,
$a_2^{(DK\pi)}/a_1^{(DK\pi)}=-0.6$~\cite{MGres}, $a_i^{(D\pi\pi)}=
a_i^{(DK\pi)},~F_0^{DK}(0)/F_0^{D\pi}(0)=1.1$, one finds using Eq.~(\ref{R2D})
the value $R_2(m_D)=0.64$ consistent with (\ref{R2number}). Still,
a probable explanation for this failure is the presence of
signficant nonfactorizable nonresonant contributions.

In view of the situation of $R_2$ at the $D$ mass, one should be aware of
the possible presence of uncalculable nonfactorizable SU(3) breaking terms at
the $B$ mass. We will disregard such terms for the rest of the discussion and
study $R_1(m_B)$ and $R_2(m_B)$ in the generalized factorization
approximation,
keeping in mind that larger SU(3) breaking may be caused by nonfactorizable
contributions.

In the factorization approximation one has
\bea\label{Rfact}
R_{1,2}(m_B) = \frac{a_{1,2}^{(BK\pi)}}{a_1^{(B\pi\pi)}+a_2^{(B\pi\pi)}}
\,\frac{f_K}{f_\pi}\,
\frac{F_0^{B\pi}(m_K^2)}{F_0^{B\pi}(m_\pi^2)} +
\frac{a_{2,1}^{(BK\pi)}}{a_1^{(B\pi\pi)}+a_2^{(B\pi\pi)}}
\,\frac{m_B^2-m_K^2}{m_B^2-m_\pi^2}\,
\frac{F_0^{BK}(m_\pi^2)}{F_0^{B\pi}(m_\pi^2)}~,
\eea
where $R_2(m_B)$ is given by an expression analogous to (\ref{R2D}).
The parameters $a_i^{(BK\pi)}$ and $a_i^{(B\pi\pi)}$ cannot be determined
direcly
from experiments. The closest one can get empirically is to measure these
parameters at a different scale, the scale of hadronic $b\to c$ decays.
An analysis of measured rates for
$B\to D^{(*)}\pi(\rho)$ and $B\to J/\psi K$, yields values \cite{BHP,NS}
$a_1^{BD\pi}\simeq 1$ and $a_2^{BD\pi}=0.2-0.3$.
A recent perturbative QCD calculation of $B\to \pi\pi$ decays \cite{BBNS},
including nonfactorizable contributions due to hard gluon exchange,
suggests that the corresponding value of the effective $a_2$ for two light
pions could be even smaller, around $a_2^{(B\pi\pi)}=0.1$.
Actually, $a_2$ acquires a sizable complex phase. This calculation does not
include nonfactorizable terms due to soft exchanges, which are argued to be
power suppressed in the heavy quark limit. A precise calculation of these soft
corrections is a challenging task. In our estimate below of the SU(3) breaking
parameters $R_{1,2}$ we will use the range $a_2=0.1 - 0.3$, assuming for
simplicity $a_i^{(BK\pi)}=a_i^{(B\pi\pi)}$. Note that in general $a_i$
acquire
complex phases and therefore $R_i$ become complex. Neglecting complex
phases has
a small effect on our estimates.

Under these assumptions, it is convenient to introduce the sum and
difference of
$R_1$ and $R_2$, in which a dependence on $(a_1-a_2)/(a_1+a_2)$ is restricted
to the difference
\bea\label{R+}
R_1 + R_2 &=& \frac{f_K}{f_\pi}\,
\frac{F_0^{B\pi}(m_K^2)}{F_0^{B\pi}(m_\pi^2)} +
\frac{m_B^2-m_K^2}{m_B^2-m_\pi^2}
\frac{F_0^{BK}(m_\pi^2)}{F_0^{B\pi}(m_\pi^2)}~,\\\label{R-}
R_1 - R_2 &=& \frac{a_1-a_2}{a_1+a_2}\left(\frac{f_K}{f_\pi}\,
\frac{F_0^{B\pi}(m_K^2)}{F_0^{B\pi}(m_\pi^2)} -
\frac{m_B^2-m_K^2}{m_B^2-m_\pi^2}\,
\frac{F_0^{BK}(m_\pi^2)}{F_0^{B\pi}(m_\pi^2)}\right)~.
\eea
The sum $R_1+R_2$ can be estimated more reliably than the difference
$R_1-R_2$, since the former does not depend on the poorly known coefficients
$a_{1,2}$.

Important ingredients entering the factorization expressions (\ref{Rfact}) are
the hadronic form factors $F_0^{BP}(q^2)$, defined in the usual way \cite{BSW}
\bea
\langle P(p_P)|\bar b\gamma_\mu q|B(p_B)\rangle =
\left( (p_B+p_P)_\mu - \frac{m_B^2-m_P^2}{q^2}q_\mu \right) F_1^{BP}(q^2) +
\frac{m_B^2-m_P^2}{q^2} q_\mu F_0^{BP}(q^2)~,
\eea
where $q=p_B-p_P$.
The form factors $F_0^{B\pi(K)}(0)$ were computed in a variety of quark
models \cite{BSW,relpotmodel}, light front model \cite{LFM}, MIT bag model
\cite{bag},
QCD sum rules \cite{BBD,QCDSR,LCSR} and lattice QCD \cite{FlSa}.
The results obtained for these form factors at $q^2=0$ are presented in
Table 2.

\begin{center}
\begin{tabular}{|l|ccccccc|}
\hline
 & BSW \cite{BSW} & QCDSR \cite{QCDSR} & LCSR \cite{LCSR} & RQM \cite{relpotmodel} &
LFM \cite{LFM} & BM \cite{bag} & Lattice QCD \cite{FlSa} \\
\hline
\hline
$F_0^{B\pi}(0)$ & 0.33 & 0.24 & $0.30\pm 0.04$  &
$0.37\pm 0.12$ & 0.26 & 0.33 & $0.27\pm 0.11$ \\
\hline
$F_0^{BK}(0)$ & 0.38 & 0.25 & $0.35\pm 0.05$ & $0.26\pm 0.08$
& 0.34 &
$-$ & $-$  \\
\hline
\end{tabular}
\end{center}
\begin{quote} {\bf Table 2.}
Theory predictions for semileptonic $B\to \pi(K)$ form
factors at $q^2=0$.
\end{quote}

The ratio of form factors $F_0^{B\pi}(m_K^2)/F_0^{B\pi}(m_\pi^2)$ is expected
to differ from 1 by less than one percent; this difference will be
neglected in the
following discussion. Using the numerical values \cite{BSW,LCSR} in Table 1
gives a typical value for the form factor ratio appearing in the second term
of (\ref{Rfact})
\beq\label{RKpi}
\frac{F_0^{BK}(m_\pi^2)}{F_0^{B\pi}(m_\pi^2)} = 1.16~.
\eeq
It is hard to assign a theoretical uncertainty to this value, considering
the large spread of model-predictions, some of which \cite{relpotmodel}
even involve values smaller than one.  The particular value (\ref{RKpi})
implies
a near cancellation of the two terms in (\ref{R-}) \cite{NR2}, giving
$R_1-R_2=0.06 (a_1-a_2)/(a_1+a_2)=0.05~(0.03)$, corresponding to
$a_2=0.1~(0.3)$. Together with the sum $R_1+R_2=2.37$, this predicts
$R_1=1.21~(1.20)$ and $R_2=1.16~(1.17)$. Thus, with the particular choice
(\ref{RKpi}), SU(3) breaking in $R_2/R_1$ is at most about $4\%$.

In view of the wide range of model-dependent results for
$F_0^{BK}(0)/F_0^{B\pi}(0)$ (see Table 2), and in order to perhaps narrow
this range, we propose an alternative
calculation of this ratio, which is based on the measured ratio of
corresponding form-factors in $D$ decays, $F_0^{DK}(0)/F_0^{D\pi}(0) =
1.00\pm 0.08$. Semileptonic $B$ and $D$ decay form factors, at points of equal
$\pi(K)$ energy in the rest frame of the decaying meson, are related by
a heavy quark symmetry scaling law \cite{IW}
\bea
F_0^{B\pi}(q_*^2) = \left(\frac{\alpha_s(m_b)}{\alpha_s(m_c)}\right)^{-6/25}
\sqrt{\frac{m_D}{m_B}}F_0^{D\pi}(0)\,,\quad
F_0^{BK}(q_*^2) =
\left(\frac{\alpha_s(m_b)}{\alpha_s(m_c)}\right)^{-6/25}
\sqrt{\frac{m_D}{m_B}}F_0^{DK}(0)~.
\eea
The momentum transfer for $B$ form factors corresponding to $q^2=0$ in $D$
decays is $q_*^2=18.0$ GeV$^2$, for $K$ in the final state, and $q_*^2=17.6$
GeV$^2$ for $\pi$.
Taking the double ratio of $B$ and $D$ form factors \cite{Grin}
cancels the leading ${\cal O}(1/m_Q)$
and ${\cal O}(m_s/\Lambda_{\chi SB})$ corrections to the scaling
laws of the individual form factors
\beq\label{doubleR}
\frac{F_0^{BK}(q_*^2)/F_0^{B\pi}(q_*^2)}
{F_0^{DK}(0)/F_0^{D\pi}(0)} = 1 + {\cal O}(m_s/m_c-m_s/m_b)~.
\eeq
We use this relation to predict the ratio of $B$ form factors (\ref{RKpi})
in terms of the corresponding ratio for $D$ decays. The extrapolation of the
former from $q_*^2$ down to $q^2=0$ is made by assuming dominance by the $0^+$
states $B_{0(s)}$ for which we take $m_{B_0}=5.7-5.8$ GeV,
$m_{B_{s0}}=5.8-5.9$
GeV.
This gives
\beq\label{RB}
\frac{F_0^{BK}(0)}{F_0^{B\pi}(0)} = (1.013\pm 0.002)
\frac{F_0^{BK}(q_*^2)}{F_0^{B\pi}(q_*^2)} \simeq 1.01\pm 0.11~,
\eeq
where we introduced an error of 7\% associated with the
${\cal O}(m_s/m_c)$ term in (\ref{doubleR}) \cite{Grin}. The rest of the
uncertainty is due to the error in $F_0^{DK}(0)/F_0^{D\pi}(0)$. This
uncertainty is expected to be reduced in future experiments of semileptonic
$D$ decays. The relation between ratios of form factors in $D$ and $B$ decays
can be tested by measuring $B\to\pi\ell\nu$ and $B\to K\ell^+\ell^-$.

The value (\ref{RB}) is somewhat lower than the result (\ref{RKpi})
taken from certain models. Inserting this value
into the relations (\ref{R+}) and (\ref{R-}), we find
$R_1 + R_2 = 2.22\pm 0.11$ and $R_1 - R_2 = (0.21 \pm 0.11)
(a_1-a_2)/(a_1+a_2)$. The central values yield
$R_1=1.20~(1.17)$ and $R_2=1.02~(1.05)$~for $a_2=0.1~(0.3)$.
This implies very small SU(3) breaking in $R_2$ and larger SU(3)
breaking in $R_2/R_1$, at a level of $15\%~(10\%)$. This is
significantly higher than the $4\%$ effect estimated from Eq.~(\ref{RKpi}).
An even larger SU(3) breaking in $R_2/R_1$ is obtained in the factorization
approximation for
values of $F_0^{BK}(0)/F_0^{B\pi}(0)$ which are smaller than 1.

We conclude with an interesting observation. Our discussion of the large
measured
SU(3) breaking in hadronic $D$ decays indicates the likely need for a
significant
nonfactorizable nonresonant contribution. Such effects may be smaller in
$B$ decays but ought to be considered with care. In spite of this warning,
one may argue from rather simple grounds that in the generalized
factorization approximation SU(3) breaking in $R_2(m_D)$ is expected to be
much larger than in $R_2(m_B)$. Assuming universal values for $a_i$,
separately
for $B$ and $D$ decays, both $R_2(m_B)$ in Eq.~(\ref{Rfact}) and $R_2(m_D)$ in
Eq.~(\ref{R2D}) consist of two SU(3) breaking contributions weighed by
$a_2/(a_1+a_2)$ and $a_1/(a_1+a_2)$. In $B$ decays, where $a_2/a_1\sim
0.1-0.3$,
the dominant $a_1$ term involves SU(3) breaking given by
$F_0^{BK}(0)/F_0^{B\pi}(0)-1$ which is expected to be at a level of $10\%$.
On the other hand, in $D$ decays in which $a_2/a_1\sim (-0.6)-(-0.4)$ is large
and negative, the $22\%$ SU(3) breaking of $f_K/f_{\pi}$ in the $a_2$ term may
be effectively roughly doubled by the destructive interference of this term
with the $a_1$ term.

\medskip

{\it Acknowledgement}: We thank H. Y. Cheng, R. Fleischer, B. Grinstein
and J. L. Rosner for useful
discussions. This work is supported in part by the National Science
Foundation,
by the United States - Israel Binational Science Foundation under Research
Grant Agreement 98-00237, and by the Israel Science Foundation
founded by the Israel Academy of Sciences and Humanities.

\def \ajp#1#2#3{Am.~J.~Phys.~{\bf#1} (#3) #2}
\def \apny#1#2#3{Ann.~Phys.~(N.Y.) {\bf#1} (#3) #2}
\def \app#1#2#3{Acta Phys.~Polonica {\bf#1} (#3) #2 }
\def \arnps#1#2#3{Ann.~Rev.~Nucl.~Part.~Sci.~{\bf#1} (#3) #2}
\def \cmp#1#2#3{Commun.~Math.~Phys.~{\bf#1} (#3) #2}
\def \cmts#1#2#3{Comments on Nucl.~Part.~Phys.~{\bf#1} (#3) #2}
\def \cn{Collaboration}
\def \corn93{{\it Lepton and Photon Interactions:  XVI International Symposium,
Ithaca, NY August 1993}, AIP Conference Proceedings No.~302, ed.~by P. Drell
and D. Rubin (AIP, New York, 1994)}
\def \cp89{{\it CP Violation,} edited by C. Jarlskog (World Scientific,
Singapore, 1989)}
\def \dpff{{\it The Fermilab Meeting -- DPF 92} (7th Meeting of the American
Physical Society Division of Particles and Fields), 10--14 November 1992,
ed. by C. H. Albright \ite~(World Scientific, Singapore, 1993)}
\def \dpf94{DPF 94 Meeting, Albuquerque, NM, Aug.~2--6, 1994}
\def \efi{Enrico Fermi Institute Report No. EFI}
\def \el#1#2#3{Europhys.~Lett.~{\bf#1} (#3) #2}
\def \epjc#1#2#3{Eur.~Phys.~J.~C~{\bf#1} (#3) #2}
\def \ibj#1#2#3{~{\bf#1} (#3) #2}
\def \ijmpa#1#2#3{Int.~J.~Mod.~Phys.~A {\bf#1} (#3) #2}
\def \ite{{\it et al.}}
\def \jhep#1#2#3{JHEP~{\bf#1} (#3) #2}
\def \jmp#1#2#3{J.~Math.~Phys.~{\bf#1} (#3) #2}
\def \jpg#1#2#3{J.~Phys.~G {\bf#1} (#3) #2}
\def \mpla#1#2#3{Mod.~Phys.~Lett.~A {\bf#1} (#3) #2}
\def \nc#1#2#3{Nuovo Cim.~{\bf#1} (#3) #2}
\def \npb#1#2#3{Nucl.~Phys.~B {\bf#1} (#3) #2}
\def \pisma#1#2#3#4{Pis'ma Zh.~Eksp.~Teor.~Fiz.~{\bf#1} (#3) #2[JETP Lett.
{\bf#1} (#3) #4]}
\def \pl#1#2#3{Phys.~Lett.~{\bf#1} (#3) #2}
\def \plb#1#2#3{Phys.~Lett.~B {\bf#1} (#3) #2}
\def \pr#1#2#3{Phys.~Rev.~{\bf#1} (#3) #2}
\def \pra#1#2#3{Phys.~Rev.~A {\bf#1} (#3) #2}
\def \prd#1#2#3{Phys.~Rev.~D {\bf#1} (#3) #2}
\def \prl#1#2#3{Phys.~Rev.~Lett.~{\bf#1} (#3) #2}
\def \prp#1#2#3{Phys.~Rep.~{\bf#1} (#3) #2}
\def \ptp#1#2#3{Prog.~Theor.~Phys.~{\bf#1} (#3) #2}
\def \rmp#1#2#3{Rev.~Mod.~Phys.~{\bf#1} (#3) #2}
\def \yaf#1#2#3#4{Yad.~Fiz.~{\bf#1} (#3) #2 [Sov.~J.~Nucl.~Phys.~{\bf #1} (#3)
#4]}
\def \zhetf#1#2#3#4#5#6{Zh.~Eksp.~Teor.~Fiz.~{\bf #1} (#3) #2 [Sov.~Phys. -
JETP {\bf #4} (#6) #5]}
\def \zpc#1#2#3{Zeit.~Phys.~C {\bf#1} (#3) #2}

\end{document}